# Two-dimensional Thyristor

Purbasha Ray[1], Rupali Verma[2], Biswajeet Nayak[1], Suman Kumar Chakraborty[1], Mayank Shrivastava[2], Prasana Kumar Sahoo[1*]

*[1]Quantum Materials and Device Research Lab, Materials Science Centre, Indian Institute of Technology Kharagpur, WB, 721302 India; [2]MSDLab, Department of Electronic Systems Engineering, Indian Institute of Science, Bangalore, 560012*

*prasana@matsc.iitkgp.ac.in

**Abstract**

Two-dimensional (2D) semiconductors offer a viable platform for next-generation electronics beyond sub-3 nm nodes, yet essential nonlinear circuit elements, such as thyristors, remain unexplored on atomically thin platforms. Here, we report an atomically thin reconfigurable electro-optical thyristor based on a lateral $WSe_2$-$MoSe_2$-$WSe_2$-$MoSe_2$ PNPN heterostructure, directly grown via sequential edge epitaxy. Unlike single-junction 2D diodes, this multijunction architecture exhibits thyristor-like nonlinear switching, including forward blocking, abrupt snapback, bistability, and electrical latching. The device exhibits a forward-blocking voltage of ~183 V across an active-material thickness of ~1.2 nm, corresponding to a critical electric field of ~1 MV $cm^{-1}$. Under open-gate conditions, avalanche-assisted carrier multiplication and regenerative feedback are associated with a sharp forward-voltage drop of 175 V, an impact-ionization index of n = 32, and a pronounced current surge and ultralow holding voltage. Gate-pulse modulation and temperature-dependent measurements indicate avalanche-assisted thyristor operation, bistable switching, and latching behavior. Notably, optical excitation reduces the turn-on voltage by approximately 70%, demonstrating optical modulation of electrical switching. The 2D lateral heterostructure, beyond diode or transistor functionality, enables a new route to nonlinear and bistable switching, latching, ultralow standby power, and adaptive logic, opening the way for reconfigurable 2D electronics and photonics ICs.

Two-dimensional (2D) semiconductors are emerging as a versatile material platform for next-generation electronics such as logic-in-memory, circuits, computing, and microprocessors [1–12], owing to their flexibility, superior electrostatic control[13], and on-demand hetero-integration [14–16]. As technologies approach sub-3 nm nodes, with increasing functional integration, device miniaturization introduces significant challenges to the reliability, protection, and adaptive functionality of integrated circuits (ICs). In conventional electronics, semiconductor-controlled rectifiers such as thyristors are essential for power electronics and circuit protection, serving as key components in high-voltage AC/DC conversion, high-voltage direct current (HVDC) systems, bistable switching, and electrostatic discharge (ESD) protection due to their high surge tolerance[17–22]. Beyond protection, thyristors offer a broad range of applications such as reconfigurable logic[23], memory and latch circuits[24], temperature and overcurrent sensing[25], optical switching, pulse power[26], and signal modulation[27], power ICs and smart drivers, highlighting their versatility across both electronic and photonic systems.

While conventional thyristors based on Si, GaAs, and SiC have facilitated robust switching in power electronics,[22,28,29] but their vertical architectures and doping constraints limit their compatibility and integration into next-generation 2D electronics and ICs. In contrast, 2D materials offer reduced dielectric screening and strong electric field confinement, which allows enhanced carrier multiplication and tunable band engineering. These properties can be modulated through vertically stacked or edge-epitaxially grown 2D lateral heterostructure (LHS), allowing efficient carrier generation and transport. In particular, 2D LHS provides drift-diffusion-limited transport across the depletion region formed at the in-plane junctions, which facilitates efficient carrier injection and voltage blocking within a planar geometry. These characteristics of 2D LHS can facilitates compact device architectures with high switching efficiency, lower power consumption, and enhanced functional integration.

Despite substantial advances in 2D transistors[30], logic gates[31], microprocessors[32], and neuromorphic devices, a circuit-level 2D ESD protection and switching element, such as a 2D thyristor, is the need of the hour, offering adaptive functionality and seamless integration across both electronic and photonic 2D circuits. Realizing a 2D analog of the conventional thyristor poses significant challenges due to the need for atomically sharp p–n–p–n junctions in a planar architecture, controlling avalanche multiplication, and stable operation under thermal and optical perturbations. Moreover, carrier multiplication and space-charge dynamics in a confined 2D architecture require careful engineering to achieve controlled and reliable switching. Consequently, atomically thin thyristors remain a critical yet missing component for advancing integrated 2D ICs in the flatland.

Here, we report the atomically thin thyristor based on a 2D $MoSe_2$-$WSe_2$ multi-junction LHS, directly grown on a CMOS-compatible substrate via one-pot edge-epitaxy[33]. The LHS-based 2D thyristor enables controlled avalanche-driven switching with regenerative feedback, exhibits bistability and high current-handling capability, and provides intrinsic ESD protection within a 1.2nm channel. This work establishes a pathway for integrating power, protection, and adaptive functionality into 2D electronic and photonic circuits, bridging materials-level advantages to system-level implementation.

The bilayer multijunction LHSs were directly grown by chemical vapor deposition (CVD) strategy on CMOS-compatible Si/$SiO_2$ substrates, exhibiting alternative $MoSe_2$ and $WSe_2$ lateral domains (Fig. 1a). Atomic-resolution high-angle annular dark-field scanning transmission electron microscopy imaging (HAADF STEM) (Fig. 1b) confirms the formation of a sharp 1D lateral interface between $MoSe_2$ and $WSe_2$ domains, with clear Z-contrast distinguishing the darker $MoSe_2$ and brighter $WSe_2$ regions arising from their atomic number difference. The lattice-resolved coherent atomic arrangement across the $MoSe_2$-$WSe_2$ interface indicates seamless epitaxial stitching and phase continuity in the $2H_C$ structure. The diffraction pattern of the selected area (Fig. 1b inset) displays well-defined diffraction spots, validating lattice coherence and high crystallinity across the heterojunction achieved through the CVD-grown LHS, essential for reliable device performance. Several 2D PNPN thyristor devices were fabricated using the as-grown $MoSe_2$-$WSe_2$ multijunction LHS on $Si^+$/$SiO_2$ substrate as the active channel material to explore the transport behavior. The 2D LHS features spatially separated in-plane domains with built-in potentials across junctions, enabling gate-tunable carrier modulation and selective carrier injection via the control-gate or optical excitation, which are critical for achieving thyristor functionality[34]. This built-in potential was confirmed by Kelvin probe force microscopy, which revealed a surface potential difference of ~ 70mV between $MoSe_2$ and $WSe_2$.

The transfer characteristics ($I_{ds}$-$V_{gs}$) for the $MoSe_2$ domain with Ni/Au contact in FET configuration exhibited n-type characteristics, while $WSe_2$ with Pd/Au contacts demonstrated p-type behavior. Both $MoSe_2$ and $WSe_2$ show linear $I_{ds}$-$V_{ds}$ output curves, indicating ohmic contact formation. The FETs achieved an $I_{on}/I_{off}$ ratio of $10^6$, confirming the electronic grade quality of the semiconducting domains. The transfer characteristics across the $MoSe_2$-$WSe_2$ junction show rectification behavior, confirming the formation of the p-n junction diode. The 2D thyristor with a lateral p–n–p–n architecture is fabricated using optimized contact engineering. The three-junction design supports bistable switching via charge carrier regeneration between the inner $n_1$ and $p_2$ layers, unlike the single-junction diode, which allows unidirectional rectification. Here, interface $J_2$ acts as a controlling junction for holding the high voltage and triggering the device to its "ON-State".[35]

Layer number-dependent thyristor characteristics reveal that bilayer devices perform better than their monolayer and trilayer counterparts. A representative *I–V* characteristic curve of the 2D thyristor is shown in Fig. 1d in the open gate condition. A positive voltage is first applied between anode (A) and cathode (K) with control gate $C_G$ left open, rendering $J_1$ ($p_1$-$n_1$) and $J_3$ ($p_2$-$n_2$) forward-biased and $J_2$ ($n_1$-$p_2$) reverse-biased.

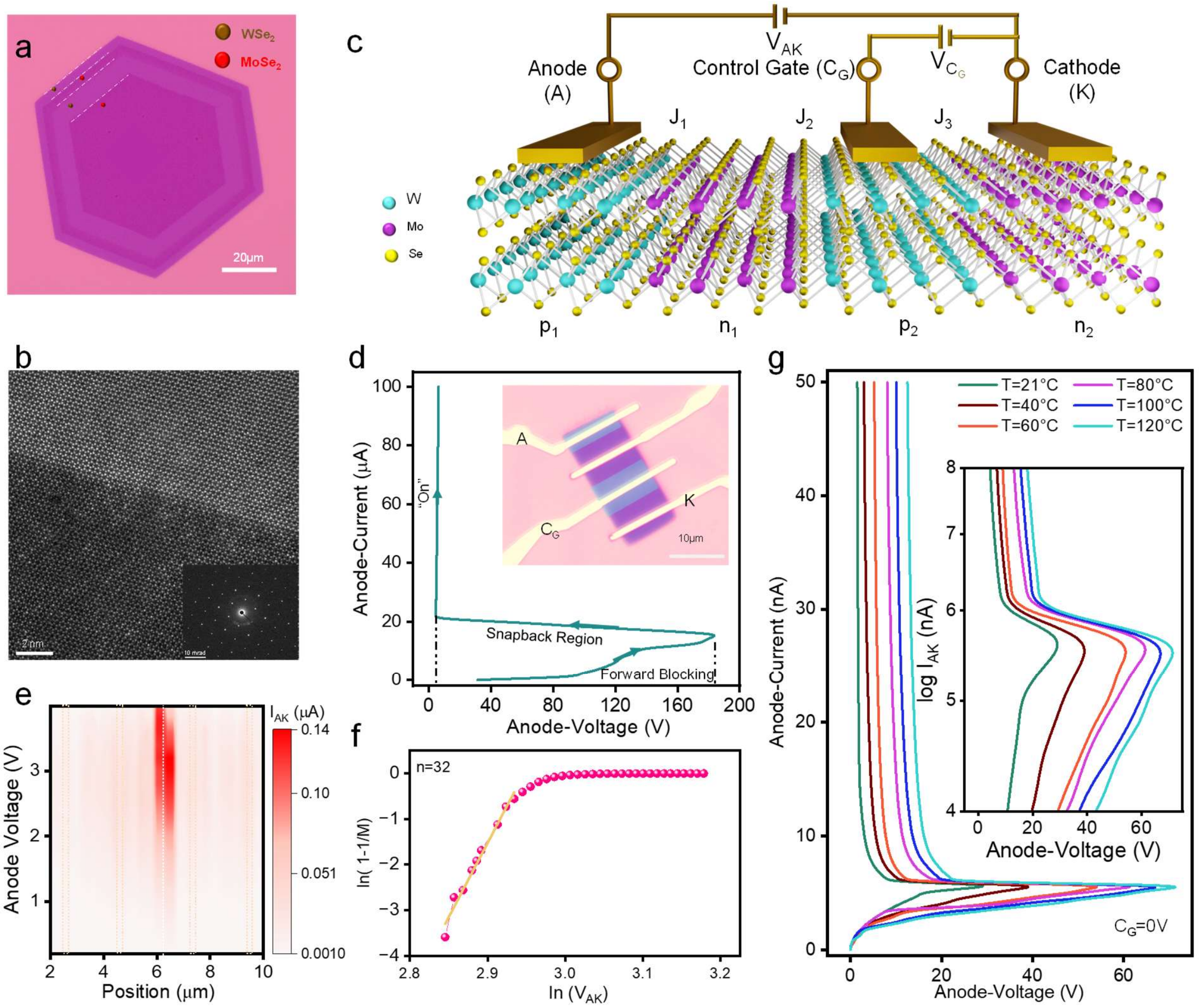

***Figure. 1**: **2D Thyristor using $MoSe_2$-$WSe_2$ Multijunction Lateral Heterostructure**. **a**. Optical micrograph of bilayer $MoSe_2$-$WSe_2$ multijunction (three junction) LHS on $SiO_2$/Si substrate, alternative contrast corresponding to $MoSe_2$ and $WSe_2$ domains. **b**. High-angle annular dark-field scanning transmission electron microscopy (HAADF-STEM) image of a bilayer $MoSe_2$-$WSe_2$ lateral interface. The inset is the selected area electron diffraction (SAED) pattern across the 2D lateral interface. **c**. Schematic representation of the 2D thyristor device. The anode (A) and cathode (C) are connected with $WSe_2$ (p-type) and $MoSe_2$ (n-type), respectively. The middle control gate (**$C_G$**) electrode on $WSe_2$ modulates the potential barrier across junctions to enable carrier transport. **d.** Anode-Current (**$I_{AK}$**) versus Anode-Voltage (**$V_{AK}$**) showing typical thyristor characteristics under current-source mode (linear scale), with a high forward breakover voltage (**$V_{BO}$**) of ~183V and a sharp voltage drop of ~175V, indicating the OFF-to-ON transition. The inset shows an optical image (false-colored $WSe_2$ domain) of the fabricated 2D thyristor device. **e**. Scanning photocurrent map using a 532 nm laser exposure (0.1 μm step size) at zero gate bias across the PNPN junctions. The maximum photocurrent is localized at the reverse-biased junction $J_2$. **f**. The avalanche mechanism at zero bias is confirmed by linearly fitting the **ln (1-1/M)** vs. **ln ($V_{AK}$)** curve (empirical model), showing the high impact ionization value n=32 for Fig. d, consistent with the efficient avalanche multiplication process. **g**. At **$C_G$**=0V, **$I_{AK}$** vs **$V_{AK}$** for different temperatures. An increase in breakdown voltage with rising temperature indicates avalanche-dominated carrier multiplication.*

The distinct snapback in the anode voltage *($V_{AK}$)* shows thyristor switching behavior with a high breakover voltage ($V_{BO}$) of ~183V (maximum) and a forward voltage drop of ~175V.The device operates in both forward blocking and conduction modes. The switching from the high-to low-impedance ("OFF" to "ON" state) is driven by a regenerative process. This transition can be triggered either by (i) avalanche breakdown at the reverse-biased $J_2$ $(n_1–p_2)$ junction (Fig. 2c) or (ii) gate-induced carrier injection, discussed subsequently. In both cases, space charge modulation at $J_2$ governs the OFF-to-ON switching mechanism.

Under open gate conditions, as $V_{AK}$ increases, the built-in electric field across $J_2$ intensifies, initiating impact ionization and subsequent avalanche breakdown, which governs current conduction in the blocking regime. The excess electrons and holes generated at $J_2$ are injected into the base of the equivalent pnp and npn transistor segments, respectively, turning both on and initiating the regenerative feedback. This feedback process turns the thyristor ON, producing the characteristic "snapback" behavior. After snapback, in the ON state, the $n_1$ and $p_2$ regions are heavily flooded with the injected electrons and holes, respectively, far exceeding their intrinsic doping concentrations and nearly eliminating the depletion region at $J_2$. The device conduction is then sustained primarily through the two forward-biased junctions (i.e., $J_1$, $J_3$). The abrupt forward voltage drops (snapback behavior) and sharp anode current increase, as indicated in the *I–V* curve, mark the transition from blocking to conduction state, a characteristic of bistable thyristor operation. The regenerative feedback underlying thyristor operation is triggered by an avalanche process localized at the high-field $J_2$ junction, corroborated by spatially resolved photocurrent mapping (Fig. 1e).

The avalanche multiplication of carriers across the junction J2 is evidenced from the exponential increase of $I_{AK}$ with photo-illumination. Meanwhile, negligible photocurrent at $J_1$ and $J_3$ confirms their forward-biased conduction state. The avalanche breakdown under open-gate conditions is further verified by fitting the *I-V* data to the empirical model[36], [$ln(1-1/M)$ vs. $ln(V_{DS})$], yielding an ionization index of $n$=32 (Fig. 1f), indicative of highly efficient avalanche multiplication. It is attributed to reduced dielectric screening, a high electric field within the narrow depletion region, and quantum confinement at the 1D interface[36,37]. At $C_G=0V$, with increasing temperature, the breakover (snapback) voltage ($V_{BO}$) increases (Fig. 1g). From the KPFM measurement, the depletion region width at $J_2$ is estimated to be ~ 1μm. The critical electric field ($E_{Critical}=V_{BO}/W_{Depletion}$) is estimated to be ~ 1MV/cm ($W_{Depletion}$: space charge region $J_2$) for the device at $C_G=0V$.

For thyristor operation, modulation of the $C_G$ voltage plays a significant role in tuning the electric field across $J_2$ junction and the band bending at $J_3$, thereby influencing the device's bistable switching characteristics. Rather than initiating the regenerative process via avalanche multiplication, the thyristor switched safely into the ON state by applying positive control gate voltage $C_G$ (Fig. 2a). Increasing $C_G$ (i) lowers the snapback voltage ($V_{BO}$), (ii) leaves the snapback current unaffected, and (iii) decreases the holding voltage ($V_H$).

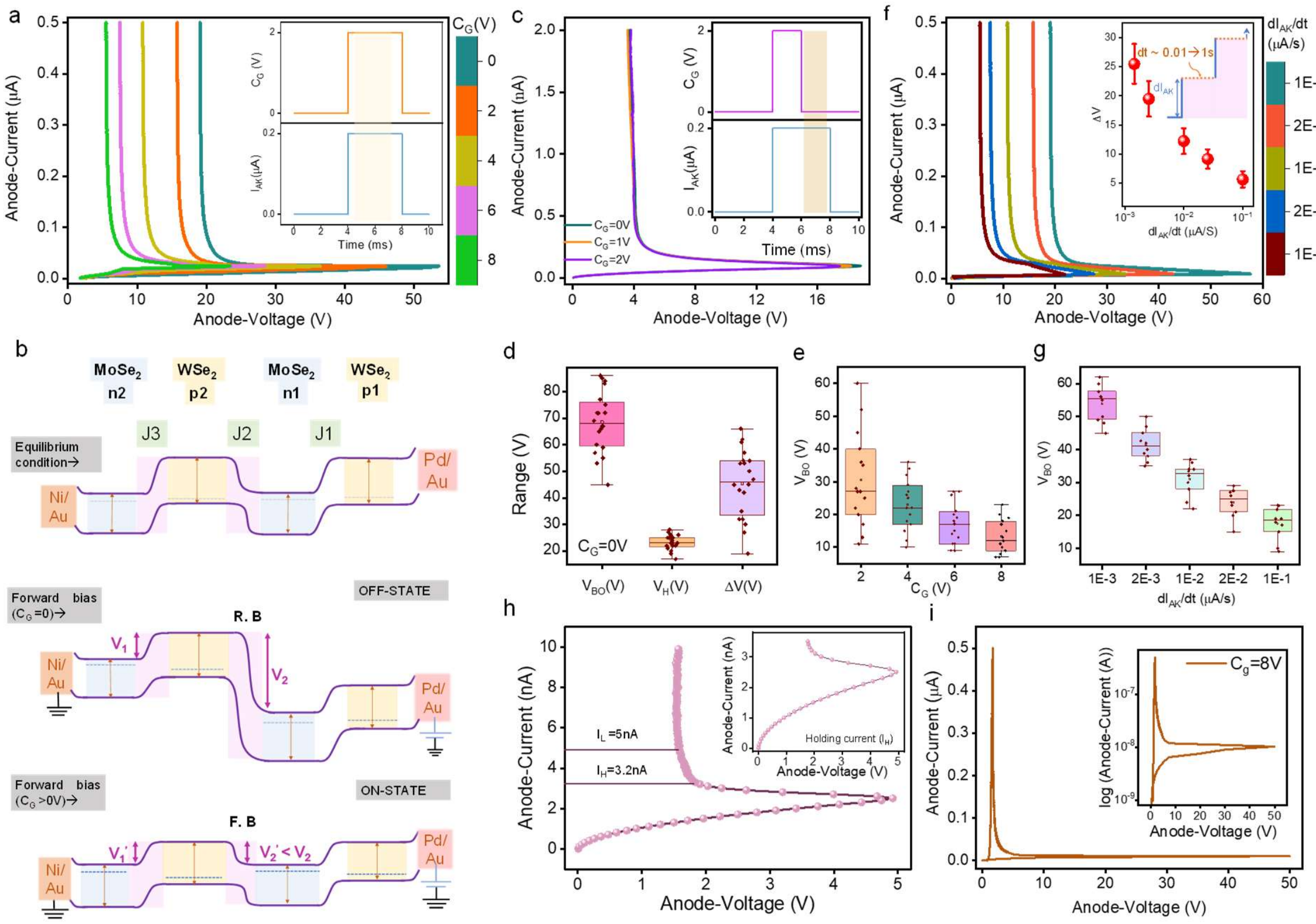

***Figure 2:*** **Electrical Characterization of Lateral Heterostructure-based 2D Thyristor**. ***a***. *Anode-Current (**$I_{AK}$**) as a function of Anode-Voltage (**$V_{AK}$**) for thyristor in current source method for different sets of $C_G$ voltage varying from 0 to 8V with a step of 2V.* ***b***. *Schematic representation of the 2D three-junction (3J) thyristor band diagram in equilibrium. Under an applied positive **$V_{AK}$**, increasing **$C_G$** voltages from zero reduces the potential barrier at $J_2$, facilitating carrier injection and enhancing current flow.* ***c***. *Linear scale pulsed gate triggered latched thyristor. Here, the holding voltage (**$V_H$**) is invariant with **$C_G$** voltage. The inset shows the Timing Diagram, indicating the pulse time used for the **$C_G$** and the corresponding $I_{AK}$ measurement to evaluate the latching and holding current. The yellow shaded region is the measurement window.* ***d.*** *At **$C_G$**=0V, the statistical plot of different devices showing three key parameters: $V_{BO}$, $V_H$, and the forward voltage drop (**ΔV**).* ***e.*** *Statistical analysis of **$V_{BO}$** for **$C_G$** varying from 2V to 8V with a step of 2V.* ***f***. *Different slew rate-based **$I_{AK}$** as a function of **$V_{AK}$**. The inset plot of **$dI_{AK}/dt$** with **ΔV** shows that with increasing **dt**, **ΔV** increases. It is controlled by constant carrier injection with time, which leads to achieving the ON state easily, resulting in an enhanced **ΔV**.* ***g***. *Statistical analysis of the **$V_{BO}$** with **$dI_{AK}/dt$** conducted for **dt** sec varying from 0.01s to 1s in steps of 0.5s. With the decreasing **dt**, the **$V_{BO}$** decreases due to faster carrier injection.* ***h.*** *A plot of both latching (**$I_L$**) and holding current (**$I_H$**) (inset fig) showing **$I_L$ > $I_H$**.* ***i.*** *Linear scale showing the hysteresis behaviour of the thyristor at **$C_G$**=8V (the inset in log scale).*

These observations can be understood from the typical energy-band diagram of a *p-n-p-n* thyristor (Fig. 2b), illustrating the modulation of the potential barriers at $J_2$ ($n_1$-$p_2$) and $J_3$ ($p_2$-$n_2$) with the applied $C_G$. Increasing $C_G$ enhances the current gain of the equivalent *n-p-n* transistor segment. Therefore, the overall gain of the equivalent *n-p-n* and *p-n-p* transistors within the 2D thyristor device is enough to assist the regenerative action at a lower $V_{AK}$, leading to an earlier snapback (lower $V_{BO}$).The observed independence of the snapback current with applied $C_G$ confirms the space-charge-modulation-induced snapback, wherein a threshold current density is required to collapse the $J_2$ space-charge region, an intrinsic device property independent of $C_G$. After snapback, the conduction is solely sustained primarily through the forward-biased junctions $J_1$ and $J_3$. A positive $C_G$ voltage also reduces the band-bending and the potential barrier across $J_3$, leading to a decrease in $V_H$ with increasing $C_G$. In contrast, in a gate-triggered latched thyristor, $V_H$ remains independent of the $C_G$ once the trigger is removed, due to the restored band-bending at $J_3$ (Fig. 2c).

The pulsed-bias timing diagram (inset, Fig. 2c) shows the measurement window (yellow shaded region) for the observation. Alternatively, injecting a gate current ($I_G$) also lowers the potential barrier at $J_2$, enabling early snapback and reducing $V_{B0}$. Conversely, applying negative gate voltage ($C_G$<0V) widens the depletion region at $J_2$ and increases the potential barrier at $J_2$ and $J_3$, thereby raising $V_{BO}$ and $V_H$. The enhanced photocurrent localized at $J_2$ under $C_G$<0V further supports this interpretation. Device-to-device variations of $V_{BO}$, $V_H$, and forward voltage drop ($\Delta V$) at $C_G$ = 0V are presented in Fig. 2d. The statistical distributions of $V_{BO}$ for $C_G$, ranging from 0 to 8V, are shown in Fig. 2e.

The dependence of the DC *I-V* characteristics on the current slew rate ($dI_{AK}/dt$) of the 2D thyristor reveals that both $V_{BO}$ and $V_H$ decrease with increasing $dI_{AK}/dt$ (Fig. 2f). When $dI_{AK}/dt$ increases rapidly, the depletion region width at junction $J_2$ fails to maintain its steady state value with time, causing transient narrowing. This narrowing of $J_2$ reduces the threshold voltage for impact ionization and regenerative feedback, triggering earlier snapback. Consequently, a higher $I_{AK}$ injection rate accelerates the collapse of $J_2$, and reduces the series resistance in the ON state, leading to smaller $\Delta V$ (inset Fig. 2f) and $V_H$. A lower *dI/dt* minimizes voltage drop, thereby reducing device heating and improving long-term stability (Fig. 2f). Statistical variation of $V_{BO}$ with $dI_{AK}/dt$ across devices is summarized in the box plot (Fig. 2g). The timing diagram in Fig. 2c (inset) illustrates the pulsed bias scheme ($C_G$: 2ms, $I_{AK}$: 4ms) used to evaluate the latching and holding currents of the 2D thyristor. To measure the latching current, $I_{AK}$ is increased until the thyristor turns ON. The device remains ON even after removing the $C_G$, establishing successful latching. Holding current can be measured by decreasing $I_{AK}$ until the thyristor turns from its ON to OFF state. As shown in Fig. 2h, the latching current exceeds the holding current, consistent with conventional thyristors. The dual (forward and then reverse)-sweep I-V hysteresis curves (Fig. 2i) further show that under $C_G$>0V, as long as a positive gate voltage is applied, a reduction in the $I_{AK}$ below the holding current prevents the device from returning to its non-conducting or blocking state. At lower temperatures, $V_{BO}$ shows a linear relationship with anode current due to space charge-limited conduction[38].

**Thermally triggered thyristor**: In 2D LHS thyristors biased in forward blocking mode at $C_G$=0V, the $V_{BO}$ increases with temperature, showing a positive temperature coefficient (Fig. 1g) due to reduced impact ionization coefficient, which decreases the multiplication coefficient. However, with $C_G$> 0V, $V_{BO}$ decreases as the thermally induced anode leakage current increases. The number of thermally generated carriers increases with increasing temperature from 25 °C to 160 °C, at $C_G = 2V$ (Fig. 3a & b), which lowers the potential barrier at $J_2$ and reduces the $V_{BO}$.

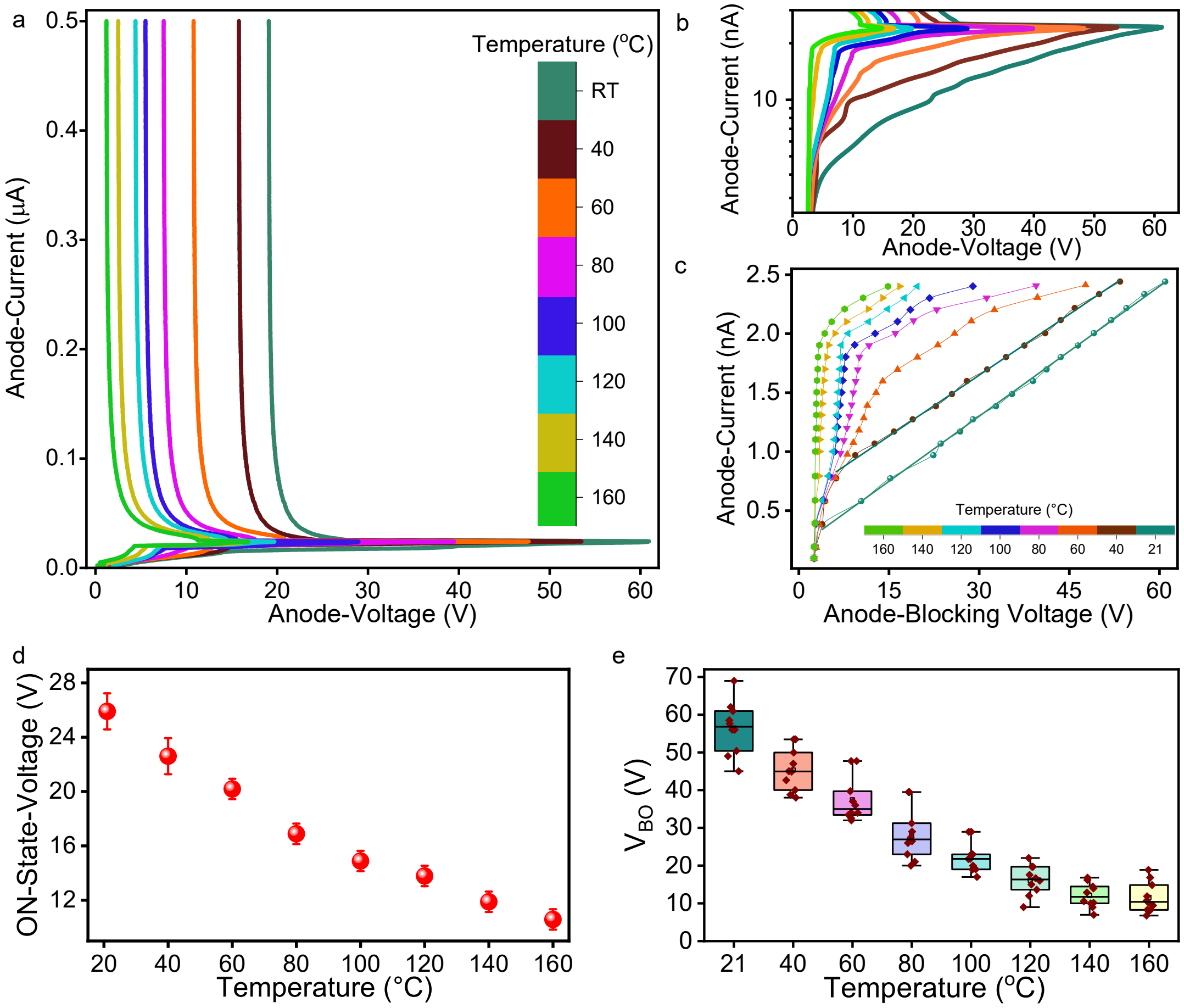

***Figure. 3.*** **Temperature-dependent 2D Thyristor Characteristics**: ***a, b***. *Linear and log scale plot of* $\boldsymbol{I_{AK}}$ *as a function of* $\boldsymbol{V_{AK}}$ *at various temperatures for a fixed* $\boldsymbol{C_G}$ *bias of +2V. With increasing the temperature, carriers gain thermal energy, reducing the potential barrier across junction* $J_2$ *and enabling faster switching from the “OFF” to “ON” state.* ***c.*** *Variation of* $\boldsymbol{I_{AK}}$ *with anode blocking voltage for different temperatures (as in panel* ***a****), exhibiting a linear behavior up to 40 °C, followed by nonlinear deviation at higher temperatures due to a decrease in potential energy.* ***d.*** *Device-to-device statistics showing the variation in forward ‘ON’ state voltage with increasing temperature, demonstrating a decreasing trend.* ***e.*** *Statistical analysis of* $\boldsymbol{V_{BO}}$ *as a function of temperature, revealing a consistent reduction. This trend indicates that elevated temperatures enhance thermally generated carrier concentrations, lowering the On-state voltage and the potential barrier at* $J_2$*, reducing the* $\boldsymbol{V_{BO}}$*.*

The rise in thermally generated carrier multiplication increases leakage current and narrows the depletion width at the reverse-biased junction $J_2$, triggering an early snapback of the thyristor. As the temperature rises, we see a deviation from the linearity due to an exponential increase in intrinsic carrier density, which leads to earlier snapback and a reduction in $V_{BO}$ (Fig. 3c). From device-to-device variability, we can see that forward ON state voltage (Fig. 3d) and $V_{BO}$ (Fig. 3e) decrease with an increase in temperature. This behavior shows a reduction in potential barriers in the 2D LHS due to thermally generated carriers. Notably, the $V_{BO}$ can be precisely tuned by temperature, making it suitable for temperature-dependent switching applications.

**Opto-thyristor Operation**: A light-triggered snapback is demonstrated in the 2D $MoSe_2$–$WSe_2$ thyristor, showing its multimodal operation capabilities. The type-II band alignment of the LHS facilitates efficient separation of photogenerated carriers. Under optical excitation, the collective effect of the strong light-matter interaction inherent to 2D TMDs, combined with the built-in field due to the space-charge region (SCR), enables efficient modulation of the potential barrier at junction $J_2$. This demonstrates that the 2D thyristor can be optically activated, which is promising in high-power electronics, optical communication, and quantum computing. Laser illumination (λ=532 nm) at the reverse-bias junction $J_2$ triggers thyristor switching under open-gate conditions (Fig. 4a). It is evidenced by snapback behavior and a significant reduction in the $V_{BO}$ compared to dark conditions in the forward-bias condition (Fig. 4b).

Illumination at the reverse-bias junction ($J_2$) with a 532 nm laser triggers the OFF-to-ON transition via avalanche photomultiplication. Unlike the avalanche-triggered snapback under dark conditions with $C_G$=0V, photoexcitation injects electron–hole pairs into the base regions of the equivalent *p-n-p* and *n-p-n* transistors, initiating a regenerative feedback mechanism and snapback. Avalanche photo-multiplication is initiated at a lower voltage than under dark conditions, as optical excitation lowers the regenerative trigger threshold voltage by photocarriers that screen the built-in electric field and narrow the depletion region, thereby enabling impact ionization at a smaller bias. Consequently, increasing optical power at $J_2$ results in a monotonic decrease in $V_{BO}$ (Fig. 4c). The forward ON-state voltage ($V$) also decreases with increasing optical power due to excess carrier injection and barrier screening (Fig. 4d). Additionally, a threefold decrease in the impact ionization coefficient is observed at 0.74μW optical power illumination at junction $J_2$. The reduction of the impact ionization index $n$ with increasing optical power indicates narrowing of the $J_2$ depletion region. A comparison of different material-based thyristor's performance metrics shows excellent forward blocking voltage with efficient avalanche mechanism at sub-nm regime. The synergistic optical and electrical triggering of the 2D thyristor establishes a reconfigurable platform for low-power and high-speed optoelectronic switching.

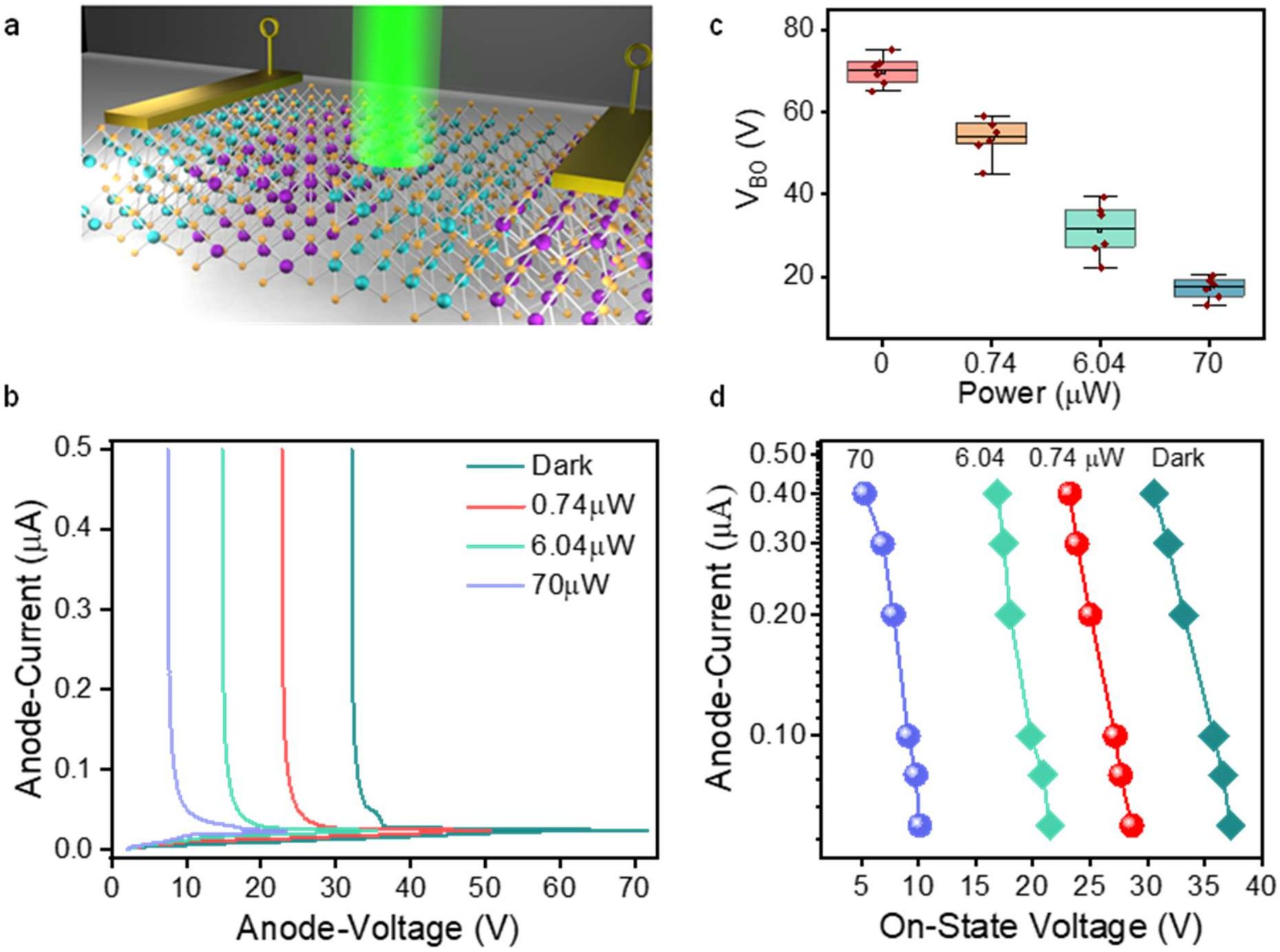

***Figure 4: Opto-thyristor based on multijunction 2D Lateral Heterostructure**: **a**. schematic illustration of the opto-thyristor configuration with a 532nm laser excitation (spot size 0.1μm) focused at the junction $J_2$ to modulate the potential barrier optically. **b**. Anode current (**$I_{AK}$**) vs. anode voltage (**$V_{AK}$**) under varying optical power, showing that with increasing exposure power, forward voltage drop (**$V_{BO}$**) reduces due to photogenerated electron-hole pairs, enabling earlier switching to the ON state. **c**. Box plot showing statistical variation in **$V_{BO}$** across multiple devices under illumination. **d**. Variation of anode current (**$I_{AK}$**) with On-State voltage (V) under illumination, demonstrating that with increasing anode current, the device switches to its low impedance state faster.*

## Conclusion

We demonstrate a gate-tunable, optically responsive, atomically thin 2D thyristor, directly grown via edge-epitaxy in a scalable, reproducible CVD process, comprising 2D LHSs of $WSe_2$-$MoSe_2$-$WSe_2$-$MoSe_2$. The device exhibits canonical thyristor behavior, including sharp snapback switching and bistability, triggered by electrical gating, thermal excitation, and optical illumination. Gate modulation, temperature-dependent transport, and spatial photocurrent mapping reveal field-driven carrier dynamics and validate the avalanche mechanism across atomically sharp 1D lateral junctions. Such 2D LHS-based thyristors show $V_{BO}$=183V and $\Delta V$=175V under open-gate conditions. A high impact ionization index (n ≈ 32) was recorded under open-gate conditions, demonstrating efficient avalanche multiplication which outperforms that of conventional semiconductors. The 2D thyristor continues to operate stably up to 160 °C, highlighting thermal resilience and electrical robustness. Thermal triggering induces a clear snapback in the 2D thyristor through the generation of thermal carriers, while the extent of forward blocking voltage depends on whether the control gate is left open or biased with a positive field. Furthermore, optical triggering accelerates snapback transition by reducing the junction barrier height and suppressing avalanche strength through photocarrier generation. Realizing a 2D opto-thyristor offers contactless, electrically isolated, fast switching.

The in-plane laterally engineered 2D heterostructures architecture on CMOS-compatible substrates offers a promising pathway for seamless integration into 2D circuits. These 2D thyristors at the atomic limit establish a viable platform for bistable switching, compact circuit protection units, and low-to-moderate power controlling devices for reconfigurable electronic and photonics ICs.

## Methods

### Direct Synthesis of 2D Lateral Multijunction Heterostructures

Monolayer, bilayer, and trilayer 2D lateral heterostructures were directly synthesized using a robust and scalable water-assisted chemical vapor deposition (CVD). Bulk $MoSe_2$ (99.99%) and $WSe_2$ (99.99%) precursors were used as sources. $Si^+/SiO_2$ (285 nm) substrates sequentially cleaned in acetone, isopropanol, and deionized water were placed within a 1-inch quartz tube furnace, ~5 cm downstream from the precursors. A mixture of source precursors ($MoSe_2$:$WSe_2$: 60:40 wt%) at zone 1 was heated to 1050 °C, while the substrates were maintained at 750-800 °C. A 200 sccm flow of ultrahigh-purity $N_2$ (99.999%) initially purged the tube for 30 min, followed by raising the temperature of Zone 1. Upon reaching 1000 °C, $N_2$ diverted through a water bubbler (4ml) was introduced to oxidize the precursors. This oxidation generated volatile transition metal oxides and chalcogen species, initiating $MoSe_2$ growth for 8 min. Abruptly switching the carrier gas to Ar+$H_2$(5%), with a flow rate of 200 sccm, suppressed the further growth of the $MoSe_2$ domain. Meanwhile, $WSe_2$ with tungsten-oxide-related precursors formed during the oxidation stage to lower sub-oxides. These preferentially deposited at $MoSe_2$ edges, enabling lateral $MoSe_2$-$WSe_2$ growth. Sequential carrier gas switching thus enabled the direct fabrication of multijunction heterostructure formation via self-aligned lateral epitaxy.

### Structural and Optical characterization

The Raman and photoluminescence (PL) Spectra were carried out using a LABRam Horiba Spectrometer with 532nm laser excitation (0.8μW) focused through a 100x objective and an 1800 grooves/mm grating. Raman and PL mappings were performed by moving the computer-controlled sample stage in the XY direction with a fixed optical path. For atomic-resolution imaging, the as-grown 2D lateral heterostructures were transferred onto TEM grids using a PMMA-assisted wet transfer process. An aberration-corrected high-angle annular dark-field scanning transmission electron microscopy (HAADF-STEM; JEOL JEM-ARM300F2) equipped with a cold-field emission gun operating at 300 kV was used. The HAADF-STEM images were acquired with a JEOL HAADF detector using a probe size of 8c and a scan speed of 20μs per pixel.

**Device Fabrication**

The grown 2D LHS samples on $Si^+/SiO_2$ substrate were patterned to define the channel regions and access individual $MoSe_2$ and $WSe_2$ domains. Initially, PMMA (495 A4/950 A2) positive-resist was spin-coated on the sample as the grown sample and patterned via electron beam lithography, followed by development in a MIBK: IPA (1:3) solution. Subsequently, oxygen plasma etching removed the exposed area within a plasma asher chamber (20 sccm $O_2$, 20W RF power, 40s). The residual resist was cleaned by dipping it in acetone for 24 hours. Further, the active channel region and electrodes were designed using e-beam lithography, followed by the deposition of metal electrodes via e-beam evaporation: Ni/Au (20/50 nm) on $MoSe_2$ and Pd/Au (10/60 nm) on $WSe_2$, selectively followed by a few lithography steps. Devices were subsequently vacuum annealed for 6 hours to improve contact quality and remove surface adsorbates.

**Electrical and Optoelectronic Measurements**

All the electrical measurements were performed using a Keithley 4200 SCS and 26XX SMUs (Keithley) in a probe station enclosed within a Faraday cage under ambient conditions. The thyristor measurements were performed in current source mode with a high input internal impedance of the SMU to prevent current surges and ensure the safe operation of the device-under-test. DC I-V characteristics were measured with current source sweeps ranging from 1nA to 100 μA in steps of ~99 nA. The temperature-dependent thyristor characteristics were conducted using a Linkam stage equipped with a PID controller feedback loop for precise thermal control during different ranges of temperatures throughout the measurement. Scanning photocurrent mappings were performed on a LABRam Horiba system with a continuous 532nm laser excitation (spot size ~ 1μm and step size ~ 0.1μm).

**Acknowledgement**

PS acknowledges the Department of Science and Technology (DST) (Project Codes: DST/NM/TUE/QM-1/2019 and National Quantum Mission (NQM) DST/QTC/NQM/QMD/2024/4/(G), India. TEM work was performed at the Sophisticated Analytical Technical Help Institute (SATHI), IIT Kharagpur, supported by DST, Govt of India. SC and BN acknowledge Mr. Arup Ghoshal for STEM measurement at SATHI. MS acknowledges NNFC (National NanoFabrication Centre) at CeNSE department, IISc for the fabrication support.

## Author Contributions

PS conceived the idea and supervised the project. PR performed the synthesis, optical characterization, device design, electro-optical measurements, and analyzed data. RV performed the device fabrication, electro-optical and KPFM measurements, and experimental design. PR and RV performed the device characterization under the supervision of MS. SC and BN contributed to CVD process development and performed STEM sample preparation and analysis. RV and BN supported the data analysis. PR and PS wrote the manuscript with RV, BN, and MS input. All authors participated in discussing the results and reviewing the manuscript.

# PR and RV contributed equally

**Additional information:** Supplementary files available online